\documentstyle[11pt,epsf,wrapfig]{article}
\begin{document}
\pagestyle{empty}
\baselineskip=0.212in

\begin{flushleft}
\Large
{SAGA-HE-110-96  
   \hfill October 31, 1996}  \\
\end{flushleft}
 
\vspace{2.0cm}
 
\begin{center}
 
\LARGE{{\bf $\bf Q^2$ evolution of structure functions}} \\
\vspace{0.3cm}

\LARGE{{\bf in the nucleon and nuclei}} \\
 
\vspace{1.3cm}
 
\LARGE
{M. Miyama $^*$ }         \\
 
\vspace{0.8cm}
  
\LARGE
{Department of Physics}         \\
 
\vspace{0.1cm}
 
\LARGE
{Saga University}      \\
 
\vspace{0.1cm}

\LARGE
{Saga 840, Japan} \\

\vspace{1.3cm}
 
\Large
{Talk given at the Circum-Pan-Pacific Workshop on} \\

\vspace{0.3cm}

{``High Energy Spin Physics'96"} \\

\vspace{0.7cm}

{Kobe, Japan, Oct. 2 -- 4, 1996 (talk on Oct. 3, 1996)}  \\
 
\end{center}
 
\vspace{1.3cm}

\vfill
 
\noindent
{\rule{6.cm}{0.2mm}} \\
 
\vspace{-0.2cm}
\normalsize
\noindent
{* Email: 96td25@cc.saga-u.ac.jp. 
   Information on his research is available}  \\

\vspace{-0.6cm}
\noindent
{at http://www.cc.saga-u.ac.jp/saga-u/riko/physics/quantum1/structure.html} \\

\vspace{1.0cm}

\vfill\eject
\pagestyle{plain}
\begin{center}
 
\Large
{\bf $\bf Q^2$ evolution of structure functions \\
in the nucleon and nuclei} \\ 
\vspace{0.7cm}
 
\large
{M. Miyama} \\
 
{Department of Physics, Saga University, Saga 840, Japan} \\
 
\vspace{1.0cm}

\normalsize
{\Large\ Abstract}
\end{center}
\vspace{-0.30cm}

$Q^2$ evolution of structure functions in the nucleon and nuclei 
is investigated by using usual DGLAP equations and 
parton-recombination equations.
Calculated results for proton's $F_2$ and for the ratio $F_2^{Ca}/F_2^D$ 
are compared with various experimental data. 
Furthermore, we study nuclear dependence of $Q^2$ evolution in tin and
carbon nuclei: $\partial [F_2^{Sn}/F_2^C]/ \partial [\ln Q^2]\ne 0$, 
which was found by NMC.

\vspace{0.6cm}

\noindent
{\Large\bf 1. Introduction}

\vspace{0.2cm}

Internal structure of the nucleon can be investigated in high-energy
lepton-nucleon scattering. Measured structure functions
depend on two variables, $Q^2=-q^2$ and $x=Q^2/2P\cdot q$, 
where $q$ is the four-momentum transfer and $P$ is the nucleon momentum.
Their $Q^2$ dependence can be calculated within perturbative QCD.
An intuitive way of describing the $Q^2$ variation
is to use the DGLAP equations \cite{DGLAP}.
They are often used in experimental analysis and
also theoretical calculations, so it is important to investigate
the numerical solution of these equations.

In addition, $Q^2$ dependence of nuclear structure functions 
becomes increasingly interesting.
It is because the NMC (New Muon Collaboration) showed 
$Q^2$ variations of the ratio $F_2^A/F_2^D$ with reasonably 
good accuracy \cite{NMCQ2}.
Furthermore, it is found recently that 
there exist significant differences between tin and carbon $Q^2$
variations, $\partial [F_2^{Sn}/F_2^C]/ \partial [\ln Q^2]\ne 0$.
It is the first indication of nuclear effects
in the $Q^2$ evolution of $F_2$ and is worth investigating
theoretically.
However, it is not obvious whether DGLAP could be applied
to the nuclear case. 
In particular, the longitudinal localization size of a parton
with momentum fraction $x$ could exceed an average nucleon
separation in a nucleus if $x$ is small ($x<0.1$).
In this case, partons in different nucleons could interact
and the interactions are called parton recombinations (PR). 
It is considered that their contributions enter into the evolution,
and modified $Q^2$ evolution equations are proposed 
in Ref. \cite{MQ}.

The purpose of our study is to investigate the numerical solution
of these $Q^2$ evolution equations.
We calculate the $Q^2$ variation of proton's $F_2$ and
the ratio $F_2^{Ca}/F_2^D$.
Nuclear dependence of $Q^2$ evolution
$\partial [F_2^{Sn}/F_2^C]/ \partial [\ln Q^2]$ 
is also calculated in order to understand the NMC measurements.

\vfill\eject

\noindent
{\Large\bf 2. Numerical solution}

\vspace{0.2cm}

The DGLAP and PR evolution equations are given by following 
integrodifferential equations:
$$
{\partial \over {\partial t}} \ q_i \left({x,t}\right)\
=\ \int_{x}^{1}{dy \over y}\ 
\left[\ \sum_j P_{q_{i} q_{j}}\left({{x \over y}}\right)\ 
           q_j \left({y,t}\right)\ 
+\  P_{qg}\left({{x \over y}}\right)\ 
g\left({y,t}\right)\ \right]
$$
$$ 
\ \ \ \ \ \ \ \ \ \ \ \ \ \ \ \ \ \ \ \ \ \ \ \ \ \ \ 
\ \ \ \ \ \ \ \ \ \ \ \ \ \ \ \ \ \ \ \ \ \ \ \ \ \ \ 
+ \ \left( recombination\ terms\ \propto \ 
{{\alpha_s A^{1/3}} \over {Q^2}} \right) \ 
\ ,
\eqno{(1a)}
$$
$$
{\partial \over {\partial t}} \ g\left({x,t}\right)\ 
=\ \int_{x}^{1}{dy \over y}\ 
\left[\ \sum_j P_{gq_j}\left({{x \over y}}\right)\ 
q_j \left({y,t}\right)\ 
+\ P_{gg}\left({{x \over y}}\right)\ 
g\left({y,t}\right)\ \right]
$$
$$
\ \ \ \ \ \ \ \ \ \ \ \ \ \ \ \ \ \ \ \ \ \ \ \ \ \ \ 
\ \ \ \ \ \ \ \ \ \ \ \ \ \ \ \ \ \ \ \ \ \ \ \ \ \ \ 
+ \ \left( recombination\ terms\ \propto \ 
{{\alpha_s A^{1/3}} \over {Q^2}} \right) \ 
\ ,
\eqno{(1b)}
$$
where the variable $t$ is defined by
$t = -(2/\beta_0) \ln [\alpha_s(Q^2)/\alpha_s(Q_0^2)]$.
The functions $q_i(x,t)$ and $g(x,t)$ are flavor-i quark 
and gluon distributions, respectively.
$P_{p_i p_j}(z)$ are called splitting functions which determine 
the probability that a parton $p_j$ with the nucleon's momentum 
fraction $y$ splits into a parton $p_i$ with the momentum fraction $x$
and another parton. 
In the PR evolution case, there are additional higher-twist terms
which are proportional to $1/Q^2$.
These terms are also proportional to $A^{1/3}$ because recombination
effect is proportional to the magnitude of parton overlap 
in the longitudinal direction.
Furthermore, there is an extra evolution
equation for a higher-dimensional gluon distribution $G_{HT}$.
Explicit expressions of recombination contributions are found 
in Ref. \cite{MQ,MK}.

As a numerical method for solving the evolution equations,
we have been studying a Laguerre-polynomial method \cite{LAG}.
It is very efficient by considering computing time and
numerical accuracy. 
However, this method has some difficulties, e.g. in handling
the non-linear recombination terms.
Therefore, we decide to employ a brute-force method \cite{MK}.
In this method, the variables $x$ and $t$ are divided into 
$N_x$ and $N_t$ steps and
integration and differentiation are defined by
$df(x)/dx=[f(x_{m+1})-f(x_m)]/\Delta x_m$ and
$\int dx f(x)=\sum_{m=1}^{N_x} \Delta x_m$ $f(x_m)$.
If initial distributions are given,
we can solve the $Q^2$ evolution equations step by step.
This is the simplest method for solving the integrodifferential
equations, but the number of steps $N_x$ and $N_t$ have to be large 
enough to get accurate results. 
Furthermore, the small $x$ region becomes increasingly important 
with the development of high-energy accelerators such as HERA.
So it is necessary to have a good accuracy at small $x$ as 
small as $10^{-5}$. In order to satisfy this condition,
the logarithmic-$x$ step 
$\Delta (log_{10} x)=|log_{10} x_{min}|/N_x$ is
taken in our analysis. 

It is useful to have a computer program for solving the evolution
equations accurately because they are frequently used in theoretical
and experimental studies.
Therefore, we provide a FORTRAN program BF1 \cite{MK} which is available 
at the following homepage

\vspace{0.3cm}
http://www.cc.saga-u.ac.jp/saga-u/riko/physics
/quantum1/structure.html, \\

\vspace{-0.2cm}
File name : bf1.fort77.gz.

\vspace{0.3cm}
\noindent
This is a very useful program for studying spin-independent structure 
functions in the nucleon and nuclei. 

\vspace{0.6cm}

\noindent
{\Large\bf 3. Results}

\vspace{0.2cm}

\begin{wrapfigure}{r}{0.46\textwidth}
  \vspace{-1.0cm}
\epsfxsize=6.2cm
\centering{\epsfbox{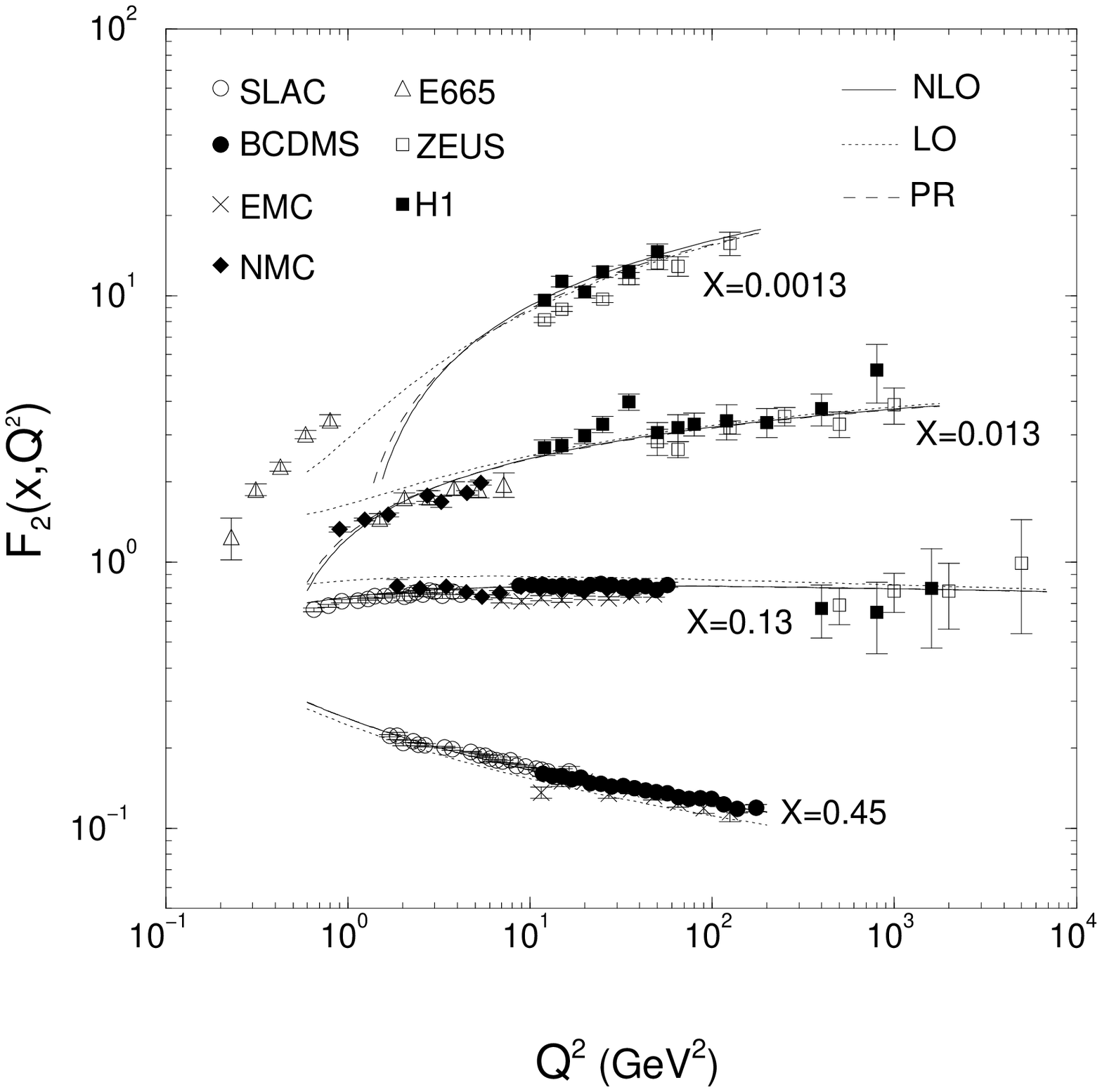}}
   \vspace{-1.9cm}
\begin{flushleft}
{Figure 1: $Q^2$ dependence of proton's $F_2$.}
\end{flushleft}
\vspace{-0.5cm}
\end{wrapfigure}

Using the program BF1, we calculate various $Q^2$ evolution.
First, we calculate $Q^2$ evolution of proton's $F_2$ structure 
function by using 
leading-order (LO) DGLAP, next-to-leading-order (NLO) DGLAP,
and parton-recombination (PR) evolution equations.
As initial distributions, we choose MRS(G) distributions
which are given at 4 GeV$^2$.
The results are shown with various experimental data in Fig. 1. 
In this figure, our results agree with the data very well 
except for the data at small $x$ and at small $Q^2$ where 
perturbative QCD would not work.
However, we can not test the recombinations 
because their effects in the nucleon are very small.

Next, we investigate nuclear cases. 
In order to calculate $Q^2$ evolution of nuclear $F_2$ structure
functions, we need to have input parton distributions in nuclei at 
certain $Q^2$. 
At this stage, there are various models \cite{F3} which can explain 
$x$ dependence of measured ratios $F_2^{A}/F_2^D$.
In such models, we have a hybrid model
with parton-recombination and $Q^2$ rescaling mechanisms \cite{SKF2,HKMU}. 
According to this model, nuclear parton distributions 
are calculated at $Q_0^2$=0.8 GeV$^2$.
We evolve these initial distributions to those at larger $Q^2$

In Fig. 2, we show the results of $F_2^{Ca}/F_2^D$ at $x$=0.0085 
with NMC experimental data \cite{NMCQ2}.
The solid, dashed, and dot-dashed curves are obtained
by LO-DGLAP, NLO-DGLAP,
and PR evolution equations respectively with $\Lambda$=0.2 GeV and $N_f$=3.
As shown by the figure, NLO and recombination
contributions to the ratio are conspicuous at small $x$.
If we evolve $F_2$ from $Q_0^2$=0.8 GeV$^2$, the recombination effect
is larger than the NLO one.
It is interesting to find such large recombination contributions.
However, it is obvious that the recombination cannot be tested 
at this stage because we do not have data at large $Q^2$ with small $x$.
In order to investigate the details of recombination, we need data
in the wide $Q^2$ region at small $x$.

\setcounter{figure}{1}
\noindent
\begin{figure}[t]
\parbox[b]{0.46\textwidth}{
  \begin{center}
    \epsfxsize=6.1cm
    \epsfbox{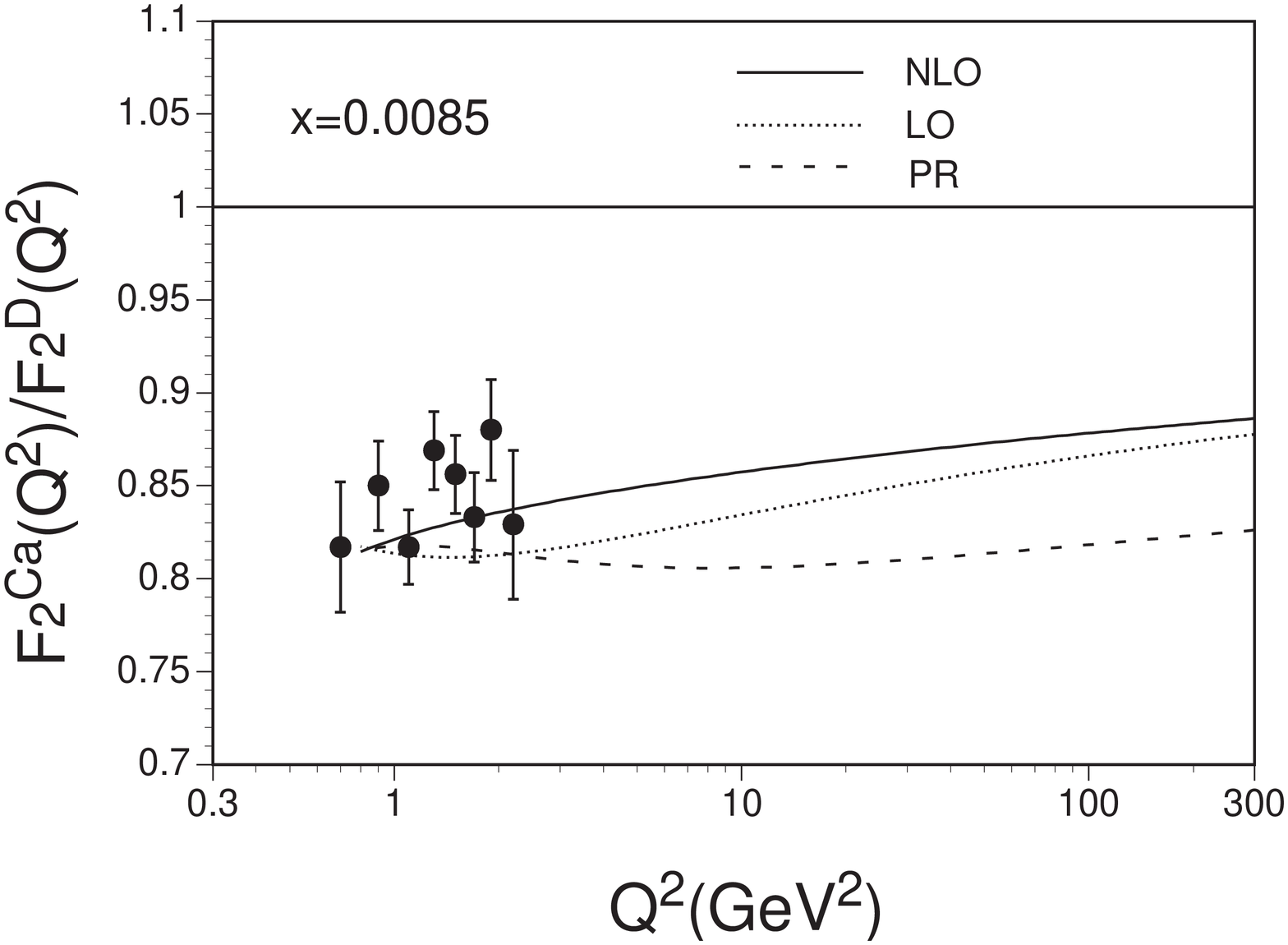}
  \end{center}
  \vspace{-0.8cm}
  \caption{$Q^2$ evolution of $F_2^{Ca}/F_2^D$.}
  \label{fig:fig2}
}\hfill
\parbox[b]{0.46\textwidth}{
  \begin{center}
    \epsfxsize=6.1cm
    \epsfbox{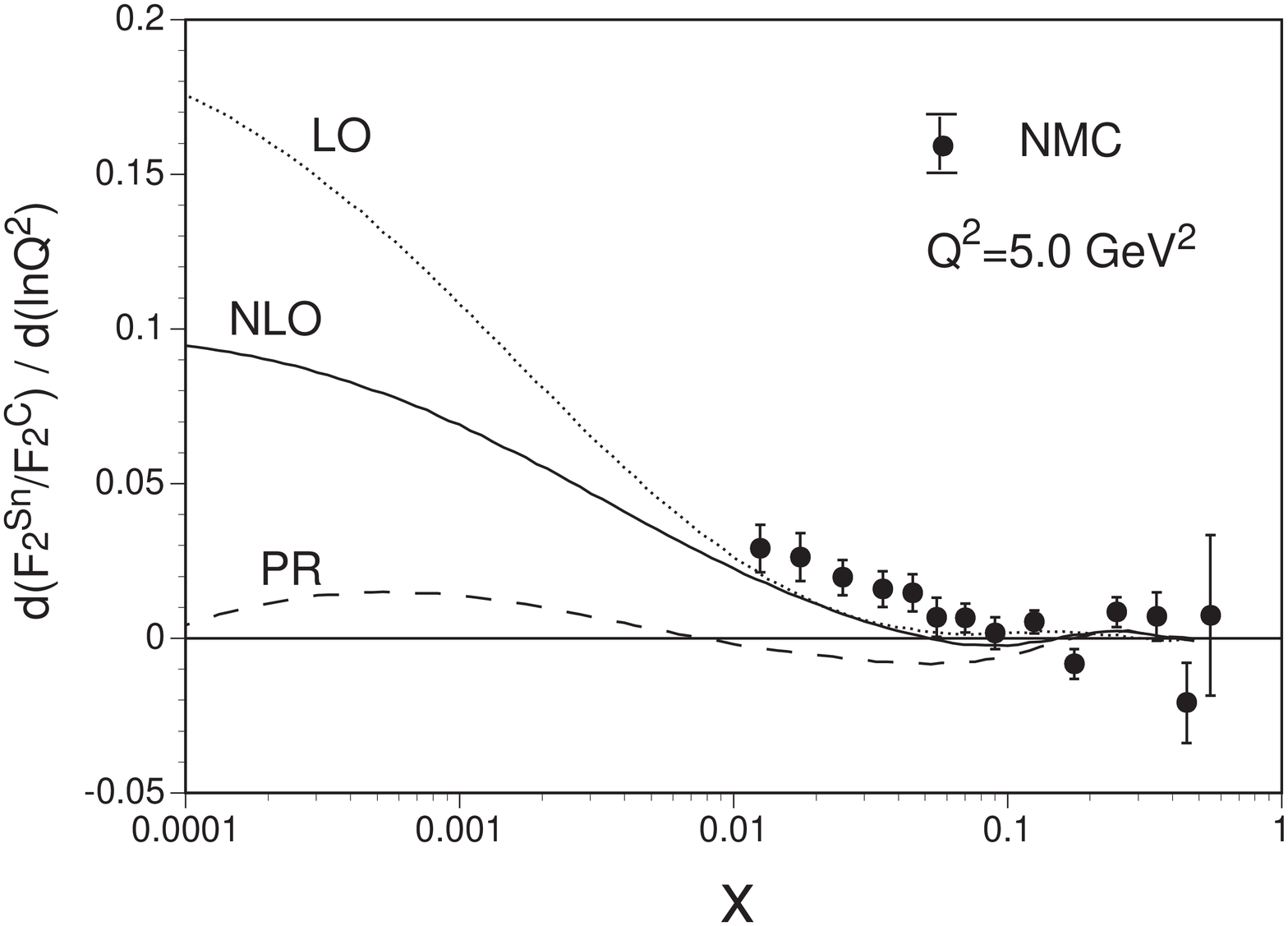}
  \end{center}
  \vspace{-0.8cm}
  \caption{Nuclear dependence of $Q^2$ evolution.}
  \label{fig:fig3}
}
\end{figure}

\vspace{-0.45cm}
We also investigate the nuclear dependence of $Q^2$ evolution in
tin and carbon nuclei \cite{KMSNC}.
Calculated results for $\partial [F_2^{Sn}/F_2^C]/ \partial [\ln Q^2]$
are shown at $Q^2$=5 GeV$^2$ together with preliminary NMC data
\cite{NMCSN-C} in Fig. 3.
The dotted, solid, and dashed curves correspond to
LO-DGLAP, NLO-DGLAP, and PR evolution results respectively.
The DGLAP evolution curves agree roughly with the experimental tendency.
It is interesting to find that the PR results disagree with
experimental data.
Because of the significant discrepancy from the data,
large parton-recombination contributions could be ruled out.
However, it does not mean that the parton-recombination mechanism
itself is in danger. 
Actually, there is an essential parameter $K_{HT}$, 
which determines how large the higher-dimensional gluon distribution is
($xG_{HT}(x,Q_0^2)=K_{HT}[xg(x,Q_0^2)]^2$).
The magnitude of $K_{HT}$ is unknown at this stage and we choose
$K_{HT}=1.68$ from Ref. \cite{MQ}.
In order to discuss the validity of the PR evolution,
the constant $K_{HT}$ must be evaluated theoretically.
It is encouraging to study the details of the
recombination mechanism in comparison with the NMC data.
On the other hand, proposed HERA nuclear program should be 
able to clarify this issue
by taking small $x$ ($\approx 10^{-4}$) data in Fig. 3.

Finally, we comment on a spin-dependent case.
Because NLO spin-dependent splitting functions are recently evaluated, 
we investigate $Q^2$ evolution of spin-dependent structure functions
with the NLO contributions.
This study is still in progress, and it is partly
discussed in Ref. \cite{HKM}.

\vfill\eject
\vspace{0.6cm}

\noindent
{\Large\bf 4. Conclusions}

\vspace{0.2cm}

We investigate numerical solution of DGLAP and PR equations.
We provide a FORTRAN program BF1 for solving these equations
in a brute-force method.
Using this program, we calculate the $Q^2$ evolution of 
proton's $F_2$ and the ratio $F_2^{Ca}/F_2^D$.
Our results are consistent with experimental data.
However, parton recombinations in the nucleon cannot be found
because their effects are very small. 
Their contributions cannot be tested at this stage even in 
the nuclear case.
In order to investigate the details of the recombination in nuclei, 
we need data in the wide $Q^2$ region at small $x$.
We also find the nuclear dependence of $Q^2$ evolution
$\partial [F_2^{Sn}/F_2^C]/ \partial [\ln Q^2]$ provides
an important clue to the recombination mechanism.

\vspace{0.6cm}

\noindent
{\Large\bf Acknowledgments} \\
\vspace{-0.2cm}

MM thanks S. Kumano for reading this manuscript.
This research was partly supported by the Grant-in-Aid for
Scientific Research from the Japanese Ministry of Education,
Science, and Culture under the contract number 06640406.
 

\end{document}